# Scalable linear solvers for sparse linear systems from large-scale numerical simulations

Hui Liu[*]and Zhangxin Chen

Department of Chemical and Petroleum Engineering

University of Calgary

Calgary, Alberta, Canada, T2N 1N4

**Abstract**

This paper presents our work on designing scalable linear solvers for large-scale reservoir simulations. The main objective is to support implementation of parallel reservoir simulators on distributed-memory parallel systems, where MPI (Message Passing Interface) is employed for communications among computation nodes. Distributed matrix and vector modules are designed, which are the base of our parallel linear systems. Commonly-used Krylov subspace linear solvers are implemented, including the restarted GMRES method, the LGMRES method, and the BiCGSTAB method. It also has an interface to a parallel algebraic multigrid solver, BoomerAMG from HYPRE. Parallel general-purpose preconditioners and special preconditioners for reservoir simulations are also developed. The numerical experiments show that our linear solvers have excellent scalability using thousands of CPU cores.

# 1 Introduction

Nowadays, various operation processes have been developed to enhance oil recovery by the oil and gas industry. Their numerical simulations are becoming more and more complicated. In the meantime, geological models from reservoirs are more and more complex, and they are also heterogenous. Models with millions of grid cells are usually employed to obtain high resolution results. Numerical simulations may take days or even longer to complete one run using regular workstations. The long simulation time could be a problem to reservoir engineers, since dozens of simulations may be required to find optimal operations. Fast computational methods and reservoir simulators should be investigated.

Reservoir simulations have been studied for decades and various models and methods have been developed by researchers, including black oil model, compositional model, thermal model and related topics. Kaarstad et al. [6] studied oil-water model and they implemented a reservoir simulator that could solve problems with up to one million grid cells. Rutledge et al. [4] developed a compositional simulator for massive SIMD computers, which employed the IMPES (implicit pressure-explicit saturation) method. Killough et al. [3] implemented a compositional simulator for distributed-memory parallel systems. Killough et al. also used the locally refined grids in their parallel simulator to improve accuracy [7]. Dogru and his group [8, 9] developed a parallel simulator, which was capable of simulating reservoir models with one billion grid cells. Zhang et al. developed a platform for adaptive finite element and adaptive finite volume methods, which has been applied to CFD, Maxwell equation, material, electronic structures, biology and reservoir simulations [10, 11, 37], and a black oil simulator using discontinuous Galerkin method has been reported [37]. For many reservoir simulations, especially black oil simulation, most of the simulation time is spent on the solution of linear systems and it is well-known that the key of accelerating linear solvers is to develop efficient preconditioners. Many preconditioner methods have been applied to reservoir simulations, including point-wise and block-wise incomplete factorization (ILU) methods for general linear systems [13], domain decomposition methods [22], constrained pressure residual (CPR) methods for the black oil model, compositional model and extended black oil models [14, 15], multi-stage methods [17], multiple level preconditioners [36] and fast auxiliary space preconditioners (FASP) [18].

[*]Authors to whom Correspondence may be addressed. Email addresses: hui.jw.liu@gmail.com

This paper presents our work on developing scalable linear solvers for large-scale reservoir simulations on parallel systems. They are implemented by C and MPI (Message Passing Interface). MPI is a standardized message-passing system designed to work on a wide varieties of parallel system and it is employed to handle communications among computation nodes. Commonly used Krylov subspace solvers are implemented, including the restarted GMRES solver and BiCGSTAB solver [12]. General preconditioners, including ILU(k), ILUT, domain decomposition [22] and AMG [32], and special preconditioners, including CPR-like preconditioners, are implemented. Detailed designs and key parameters are presented. Numerical experiments show that our linear solvers are efficient and capable of calculating linear systems with hundreds of millions of unknowns They also have excellent scalability on distributed-memory parallel computers.

## 2 Distributed Matrix and Vector

A linear system, $Ax = b$ ($A \in R^{N \times N}$), is solved, where matrix and vectors are distributed among all processors. This section introduces distributed-memory matrix, `MAT`, vector, `VEC` and map, `MAP`.

### 2.1 Map

In our design, each row is owned by one MPI process, and each row of a distributed matrix has a unique global row index, which ranges from 0 to $N-1$ consecutively and is numbered from the 1st MPI process to the $N_p$-th MPI process. A vector is distributed the same way. Each global index also has a local index, which starts from 0 in each MPI process.

`MAP` is designed to management their distributions, whose data structure is shown by

Figure 1. `L2G` defines local index to global index mapping. `offset`, `ilower` and `iupper` define global index distribution. `nlocal` is the number of rows of a sub-matrix in current process. `nglobal` is the number of rows of the matrix. For any MPI process $i$, the following conditions are satisfied,

$$\begin{cases} \texttt{offset[i]} = \texttt{ilower} \\ \texttt{offset[i + 1]} = \texttt{iupper} \\ \texttt{offset[i]} \leq j < \texttt{offset[i + 1]}, \texttt{j is any global index in current process} \\ \texttt{offset[0]} = 0 \\ \texttt{nlocal} = \texttt{offset[i + 1]} - \texttt{offset[i]} \\ \texttt{offset}[N_p] = \texttt{nglobal} \end{cases} \quad (1)$$

```
typedef struct MAP_
{
    INT         *L2G;          /* local index to global index */

    INT         *offset;
    INT         ilower;
    INT         iupper;

    INT         nlocal;
    INT         ntotal;
    INT         nglobal;

    MPI_Comm    comm;
    int         rank;
    int         nprocs;

    BOOLEAN     assembled;

} MAP;
```

Figure 1: Data structure of mapping

### 2.1.1 Create

Figure 2 shows how to create a map. `comm` is the MPI communicator. `ilower` and `iupper` define global index range, where an index $j$ must satisfy

$$\texttt{ilower} \leq j < \texttt{iupper}, \texttt{ilower} < \texttt{iupper} \quad (2)$$

```
MAP * MapCreate(INT ilower, INT iupper, MPI_Comm comm);
```

Figure 2: (A) Declaration of map creating function

Another way to create a map is shown by Figure 3, where `nglobal`, `ilower`, `iupper` and other members can be calculated.

```
MAP * MapCreate(INT nlocal, MPI_Comm comm);
```

Figure 3: (B) Declaration of map creating function

### 2.1.2 Destroy

The following function is used to destroy a map, including freeing memory.

```
void MapDestroy(MAP *map);
```

Figure 4: Map destroy function

### 2.1.3 Assemble

Each map defines global index distribution and local index numbering. Local indices include indices that belong to current MPI process and indices belong to other MPI processes. Therefore, local indices should be calculated. The following function is used to assemble a map, including allocating memory, calculating local index and assembling communication information.

```
void MapAssembleByMat(MAP *map, MAT *A);
```

Figure 5: Map assemble function

## 2.2 Vector

A distributed floating-point vector is defined in Figure 6, which has buffer that holds data entries (`data`), number of local entries belong to current MPI task (`nlocal`), and number of total entries (including off-process entries, `ntotal`). Usually, `ntotal` is larger than `nlocal`. The length of `data` is equal to `ntotal`.

```
typedef struct VEC_
{
    FLOAT   *data;
    MAP     *map;
    INT     nlocal;   /* entries belong to current proc */
    INT     ntotal;   /* total entries in current MPI process */

} VEC;
```

Figure 6: Data structure of VEC

### 2.2.1 Vector Operations

Basic operations are shown by Figure 7 to Figure 13, including creating, destroying, adding entry, getting entry, and basic algebraic operations.

```
VEC * VecCreate(MAP *map);
```

Figure 7: Vector create

```
void VecDestroy(VEC **vec);
```

Figure 8: Vector destroy

```
/* index is local index */
void VecAddEntry(VEC *vec, INT idx, FLOAT value);

/* index is local index */
FLOAT VecGetEntry(VEC *vec, INT idx);
```

Figure 9: Vector add and get entry

```
/* y = a * x + b * y */
void VecAXPBY(FLOAT alpha, VEC *x, FLOAT beta, VEC *y)
{
    INT i;

    assert(x != NULL && y != NULL);
    assert(x->nlocal == y->nlocal);

#if USE_OMP
#pragma omp parallel for
#endif
    for (i = 0; i < x->nlocal; i++) {
        y->data[i] = x->data[i] * alpha + beta * y->data[i];
    }
}
```

Figure 10: y = a * x + b * y

```
/* z = a * x + b * y */
void VecAXPBYZ(FLOAT alpha, VEC *x, FLOAT beta, VEC *y, VEC *z)
{
    INT i;

    assert(x != NULL && y != NULL && z != NULL);
    assert(x->nlocal == y->nlocal && z->nlocal == y->nlocal);

#if USE_OMP
#pragma omp parallel for
#endif
    for (i = 0; i < x->nlocal; i++) {
        z->data[i] = x->data[i] * alpha + beta * y->data[i];
    }
}
```

Figure 11: z = a * x + b * y

```
/* d = <x, y> */
FLOAT VecDot(VEC *x, VEC *y)
{
    INT i;
    FLOAT s, t;

    assert(x != NULL && y != NULL);
    assert(x->nlocal == y->nlocal);

    s = 0.;
    for (i = 0; i < x->nlocal; i++) s += x->data[i] * y->data[i];

    MPI_Allreduce(&s, &t, 1, SOLVER_MPI_FLOAT, MPI_SUM, x->map->comm);

    return t;
}
```

Figure 12: `d = <x, y>`

```
/* set value */
void VecSetValue(VEC *x, FLOAT v)
{
    INT i;

    assert(x != NULL);

#if USE_OMP
#pragma omp parallel for
#endif
    for (i = 0; i < x->nlocal; i++) x->data[i] = v;
}
```

Figure 13: Set value

## 2.3 Matrix

Its distribution is demonstrated by Figure 14, which shows a matrix distributes in four MPI processes. Each MPI process own a sub-matrix and the matrix is divided by row.

When a sub-matrix is stored in some process, each row also has a local index in each MPI process that starts from 0. Each column also has a local index. The global indices of a vector and its local indices are numbered the same way. A mapping is important, which can be used to describe matrix distribution.

The data structure of a distributed matrix is more complex than a vector, which requires entries for each row and some other additional information. It is represented in Figure 15. The `ROW` stores non-zero entries of each row and its storage format is similar to a CSR matrix, which has a value of an entry (`data`), the global index (`gcol`) local index of an entry (`lcols`), and number of entries (`ncols`).

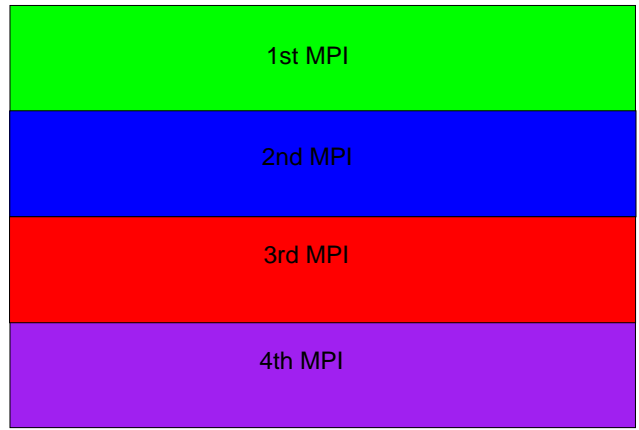

Figure 14: Matrix distribution and each MPI owns a sub-matrix.

```
/* struct for a matrix row */
typedef struct MAT_ROW_
{
    FLOAT    *data;     /* data */
    INT      *lcols;    /* local column indices, INT[ncols] */
    INT      *gcols;    /* global column indices, INT[ncols] */
    INT      ncols;     /* number of nonzero columns */
    INT      alloc;

} MAT_ROW;

typedef struct MAT_
{
    MAT_ROW     *rows;
    MAP         *map;

    INT         nlocal;     /* local entries belong to current proc */
    INT         ntotal;     /* number of columns with entries */
    INT         nglobal;    /* global matrix size */

    int         rank;
    int         nprocs;
    MPI_Comm    comm;
    BOOLEAN     assembled;

} MAT;
```

Figure 15: Data structure of MAT

### 2.3.1 Matrix-Vector Operations

Figure 16 to Figure 21 show creating, destroying, assembling, and the most important matrix-vector operations, including $y = A * x$ and $y = a * A * x + b * y$.

```
MAT * MatCreate(MAP *map);
```

Figure 16: Matrix create

```
void MatDestroy(MAT **mat);
```

Figure 17: Matrix destroy

```
/* row: local index, gcols: global indices */

void MatAddEntries(MAT *mat, INT row, INT ncols, INT *gcols, FLOAT *values);
```

Figure 18: Matrix destroy

```
void MatAssemble(MAT *mat);
```

Figure 19: Matrix assemble

```
/* y = Ax */
void MatVecMXY(MAT *A, VEC *x, VEC *y)
{
    INT i;

    assert(A != NULL && x != NULL && y != NULL);
    assert(A->nlocal == x->nlocal && x->nlocal == y->nlocal);
    assert(y->nlocal >= 0);

    /* gather off-process vector */
    MatVecGatherVec(x, A);

#if USE_OMP
#pragma omp parallel for
#endif
    for (i = 0; i < y->nlocal; i++) {
        FLOAT t;
        INT j;

        t = 0.;
        for (j = 0; j < A->rows[i].ncols; j++) {
            t += A->rows[i].data[j] * x->data[A->rows[i].lcols[j]];
        }

        y->data[i] = t;
    }
}
```

Figure 20: y = A * x

# 3 Linear Solvers

For the linear system, $Ax = b$, derived from a nonlinear method, Krylov subspace solvers including the restarted GMRES(m) solver, the BiCGSTAB solver, and algebraic multi-grid

```c
/* y = a * Ax + b * y */
void MatVecAMXPBY(FLOAT alpha, MAT *A, VEC *x, FLOAT beta, VEC *y)
{
    INT i;

    assert(A != NULL && x != NULL && y != NULL);
    assert(A->nlocal == x->nlocal && x->nlocal == y->nlocal && y->nlocal >= 0);

    /* gather off-process vector */
    MatVecGatherVec(x, A);

#if USE_OMP
#pragma omp parallel for
#endif
    for (i = 0; i < y->nlocal; i++) {
        FLOAT t;
        INT j;

        t = 0.;
        for (j = 0; j < A->rows[i].ncols; j++) {
            t += A->rows[i].data[j] * x->data[A->rows[i].lcols[j]];
        }

        y->data[i] = alpha * t + beta * y->data[i];
    }
}
```

Figure 21: y = a * A * x + b * y

(AMG) solvers are commonly used to find its solution. The Krylov subspace solvers mentioned here are suitable for arbitrary linear systems while the algebraic multi-grid solvers are efficient for positive definite linear systems.

Commonly used Krylov methods are implemented, including the restarted GMRES method, the LGMRES method, the Conjugate Gradient method, and the BiCGSTAB method. The GMRES method and the Conjugate Gradient method are described by Algorithm 1 and 2.

The data structure of our solvers, `SOLVER`, is shown in Figure 22, which includes parameters (`rtol`, `atol`, `btol`, `maxit`, `restart`), matrices, right-hand sides, solutions, and preconditioner information.

## 3.1 Solver Operations

Figure 23 to Figure 28 show solver management functions, including creating, destroying, assembling, entry adding and solver settings.

# 4 Preconditioners

Several preconditioners are developed, including general purpose preconditioners and physics-based preconditioners for reservoir simulations only.

**Algorithm 1** The preconditioned GMRES(m) algorithm
1: **for** j = 1, 2, $\cdots$ **do**
2:     Solve $r$ from $Mr = b - Ax^0$
3:     $v^1 = r/\|r\|_2$
4:     $s = \|r\|_2 e_1$
5:     **for** i = 1, 2, $\cdots$, m **do**
6:         Solve $w$ from $Mw = Av^i$
7:         **for** k = 1, $\cdots$, i **do**
8:             $h_{k,i} = \langle w, v^k \rangle$
9:             $w = w - h_{k,i} v^k$
10:         **end for**
11:         $h_{i+1,i} = \|w\|_2$
12:         $v^{i+1} = w/h_{i+1,i}$
13:     **end for**
14:     Compute vector $y$ which minimizes $\|s - H_m y\|$ and $x^m = x^0 + y_1 \times v^1 + \cdots + y_m \times v^m$
15:     If satisfied then stop, else set $x^0 = x^m$
16: **end for**

**Algorithm 2** Conjugate Gradient algorithm
1: $\vec{r} = \vec{b} - A\vec{x_0}$; $\vec{x_0}$ is an initial guess vector
2:                                                                                    ▷ SpMV; Vector update
3: **for** k = 1, 2, $\cdots$ **do**
4:     Solve $\vec{z}$ from $M\vec{z} = \vec{r}$     ▷ Preconditioner system
5:     $\rho_{k-1} = (\vec{r}, \vec{z})$     ▷ Dot product
6:     **if** $k = 1$ **then**
7:         $\vec{p} = \vec{z}$
8:     **else**
9:         $\beta = (\rho_{k-1}/\rho_{k-2})$
10:         $\vec{p} = \vec{z} + \beta\vec{p}$     ▷ Vector update
11:     **end if**
12:     $\vec{q} = A\vec{p}$     ▷ SpMV
13:     $\alpha = \rho_{k-1}/(\vec{p}, \vec{q})$     ▷ Dot product
14:     $\vec{x} = \vec{x} + \alpha\vec{p}$     ▷ Vector update
15:     $\vec{r} = \vec{r} - \alpha\vec{q}$     ▷ Vector update
16:     **if** $\|\vec{r}\|_2$ is satisfied **then**     ▷ Dot product
17:         Stop
18:     **end if**
19: **end for**

The data structure for preconditioners is defined by Figure 29. It provides three function pointers that can complete assembling (`PC_SETUP`), solving (`PC_SOLVE`) and destroying (`PC_DESTRORY`) a preconditioning system. With these function pointers, this data structure is general purpose, and if a new set of implementations are provided, a new preconditioner can be constructed.

```
typedef struct SOLVER_
{
   /* return values */
   FLOAT              residual;
   int                nits;

   /* parameters */
   FLOAT              rtol;
   FLOAT              atol;
   FLOAT              btol;
   INT                maxit;
   INT                aug_k;
   INT                restart;

   MAT                *A;
   VEC                *rhs;
   VEC                *x;

   /* preconditioner */
   SOLVER_PC          pc;
   int                pc_type;

   /* MPI info */
   int                rank;
   int                nprocs;
   MPI_Comm           comm;

   BOOLEAN            assembled;

} SOLVER;
```

Figure 22: Data structure of SOLVER

```
SOLVER * SolverCreate(SOLVER_TYPE solver_type, PC_TYPE pc_type, MAP *map);
```

Figure 23: Solver create

```
int SolverDestroy(SOLVER **solver_ptr);
```

Figure 24: Solver destroy

```
int SolverAssemble(SOLVER *solver);
```

Figure 25: Solver assemble

```
int SolverSolve(SOLVER *solver, VEC *x);
```

Figure 26: Solver solve

## 4.1 RAS Method

For the classical ILU methods, the given matrix $A$ is factorized into a lower triangular matrix $L$ and an upper triangular matrix $U$; a lower triangular linear system and an upper triangular

```
void SolverAddMatEntry(SOLVER *solver, INT row, INT gcol, FLOAT value);
void SolverAddMatEntriesByMat(SOLVER *solver, MAT *A);
void SolverAddRHSEntry(SOLVER *solver, INT idx, FLOAT value);
```

Figure 27: Solver add entry

```
/* settings */
void SolverSetDefaultMaxit(INT it);
void SolverSetDefaultRtol(FLOAT tol);
void SolverSetDefaultAtol(FLOAT tol);
void SolverSetDefaultBtol(FLOAT tol);
void SolverSetDefaultRestart(INT m);
void SolverSetDefaultAugk(INT k);

void SolverSetMaxit(SOLVER *s, INT it);
void SolverSetRtol(SOLVER *s, FLOAT tol);
void SolverSetAtol(SOLVER *s, FLOAT tol);
void SolverSetBtol(SOLVER *s, FLOAT tol);
void SolverSetRestart(SOLVER *s, INT m);
void SolverSetAugk(SOLVER *s, INT k);
```

Figure 28: Solver settings

```
/* preconditioner interface */
typedef void (*PC_SETUP)(struct SOLVER_PC_ *pc, MAT *mat);
typedef void (*PC_SOLVE)(struct SOLVER_PC_ *pc, VEC *x, VEC *b);
typedef void (*PC_DESTROY)(struct SOLVER_PC_ *pc);

/* SOLVER_PC struct */
typedef struct SOLVER_PC_
{
    struct SOLVER_      *solver;     /* pointer to solver */

    void                *data;
    PC_SETUP            setup;
    PC_SOLVE            solve;
    PC_DESTROY          destroy;

    /* MPI */
    int                 rank;
    int                 nprocs;
    MPI_Comm            comm;

    BOOLEAN             assembled;

} SOLVER_PC;
```

Figure 29: Data structure of preconditioners

linear system are required to solve:

$$\mathtt{LU\ x\ =\ b} \iff \mathtt{Ly\ =\ b,\ Ux\ =\ y}. \tag{3}$$

The systems need to be solved row-by-row, which are serial. It is well-known that they have limited scalability. Another option for parallel computing is the restricted additive Schwarz (RAS) method [20], which was developed by Cai et al.

```
typedef struct RAS_PARS_
{
    INT    overlap;
    INT    iluk_level;
    INT    ilut_p;
    int    solver;
    FLOAT  ilut_tol;
    FLOAT  filter_tol;
    INT    ilutc_drop;

} RAS_PARS;

/* RAS */
typedef struct RAS_DATA_
{
    mat_csr_t L;
    mat_csr_t U;

    FLOAT     *frbuf;
    FLOAT     *fxbuf;
    FLOAT     *fbbuf;
    INT       *ras_pro;     /* ras prolongation */
    INT       num_ras_pro;

    RAS_PARS  pars;

} RAS_DATA;
```

Figure 30: Data structure of RAS preconditioner

```
static RAS_PARS ras_pars_default =
{
    /* overlap */       1,
    /* k */             0,
    /* ilut_p */        -1,
    /* solver */        ILU(1),
    /* ilut_tol */      1e-3,
    /* filter tol */    1e-4,
    /* drop */          0,
};
```

Figure 31: Default parameters of RAS preconditioner

The data structure of the RAS preconditioner is shown in Figure 30. The `pars` stores parameters of the RAS preconditioner, such as overlap, local solver (ILUK, ILUT and ILTC), the level of ILUK, memory control and tolerance of ILUT, and filter tolerance. The `RAS_DATA` has a local problem stored by the lower triangular matrix `L` and the upper triangular matrix `U`, memory buffer and prolongation (restriction) operation information.

The default parameters of the RAS preconditioner is shown in Figure 31. Its default local solver is ILU(1). Default parameters for ILUT(p, tol) is -1 and 1e-3. If $p$ is -1, the factorization subroutine will determine dynamically.

## 4.2 Algebraic Multi-Grid (AMG) Method

If $A$ is a positive-definite square matrix, the AMG methods [30, 29, 26, 27, 28, 2] have proved to be efficient methods and they are also scalable [21]. AMG methods have hierarchical structures, and a coarse grid is chosen when entering a coarser level. Its structure of an algebraic multigrid solver is shown in Figure 32.

A restriction operator $R_l$ and an interpolation (prolongation) operator $P_l$ are determined. In general, the restriction operator $R_l$ is the transpose of the interpolation (prolongation) operator $P_l$:

$$R_l = P_l^T.$$

The matrix on the coarser grid is calculated by

$$A_{l+1} = R_l A_l P_l. \qquad (4)$$

We know that a high frequency error is easier to converge on a fine grid than a low frequency error, and for the AMG methods, the restriction operator, $R_l$, projects the error from a finer grid onto a coarser grid and converts a low frequency error to a high frequency error. The interpolation operator transfers a solution on a coarser grid to that on a finer grid. Its setup phase for the AMG methods on each level $l$ ($0 \leq l < L$) is formulated in Algorithm 3, where a coarser grid, an interpolation operator, a restriction operator, a coarser matrix and post- and pre-smoothers are constructed. By repeating the algorithm, an $L$-level system can be built. The solution of the AMG methods is recursive and is formulated in Algorithm 4, which shows one iteration of AMG. The AMG package we use is the BoomerAMG from HYPRE [32].

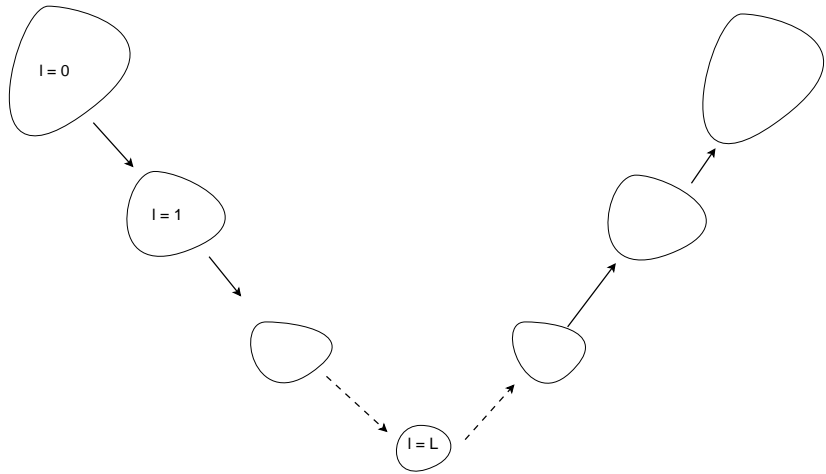

Figure 32: Structure of AMG solver.

**Algorithm 3** AMG setup Algorithm
1: Calculate strength matrix $S$.
2: Choose coarsening nodes set $\omega_{l+1}$ according to strength matrix $S$, such that $\omega_{l+1} \subset \omega_l$.
3: Calculate prolongation operator $P_l$.
4: Derive restriction operator $R_l = P_l^T$.
5: Calculate coarse matrix $A_{l+1}$: $A_{l+1} = R_l \times A_l \times P_l$.
6: Setup pre-smoother $S_l$ and post-smoother $T_l$.

**Algorithm 4** AMG V-cycle Solution Algorithm: amg_solve($l$)
Require: $b_l$, $x_l$, $A_l$, $R_l$, $P_l$, $S_l$, $T_l$, $0 \leq l < L$

$b_0 = b$
**if** ($l < L$) **then**
    $x_l = S_l(x_l, A_l, b_l)$
    $r = b_l - A_l x_l$
    $b_{r+1} = R_l r$
    amg_solve($l+1$)
    $x_l = x_l + P_l x_{l+1}$
    $x_l = T_l(x_l, A_l, b_l)$
**else**
    $x_l = A_l^{-1} b_l$
**end if**
x = $x_0$

The Cleary-Luby-Jones-Plassman (CLJP) parallel coarsening algorithm was proposed by Cleary [23] based on the algorithms developed by Luby [25] and Jones and Plassman [24]. The standard RS coarsening algorithm has also been parallelized [32]. Falgout et al. developed a parallel coarsening algorithm, the Falgout coarsening algorithm, which has been implemented in HYPRE [32]. Yang et al. proposed HMIS and PMIS coarsening algorithms for a coarse grid selection [1]. Various parallel smoothers and interpolation operators have also been studied by Yang et al [32, 1].

The AMG solvers are from HYPRE. They can be used as solvers and preconditioners. The data structure of the AMG solvers is shown in Figure 33. The `BMAMG_PARS` stores parameters of the AMG method, including the coarsening type, interpolation type, maximal levels, smoother type, and cycle type. The `BMAMG_DATA` stores linear system information, such as a distribution pattern of matrices and vectors, mapping information, and related interfaces.

Default parameters for the AMG method is shown in Figure 34, where a default six-level AMG method is applied. The detailed explanation of each parameter can be read from the HYPRE user manual.

```
typedef struct BMAMG_PARS_
{
    INT         maxit;
    INT         num_funcs;          /* size of the system of PDEs */
    INT         max_levels;         /* max MG levels */
    FLOAT       strength;           /* strength threshold */
    FLOAT       max_row_sum;        /* max row sum */
    FLOAT       trunc_tol;          /* trunc tol */

    int         coarsen_type;       /* default coarsening type = Falgout */
    int         cycle_type;         /* MG cycle type */
    int         relax_type;         /* relaxation type */
    int         coarsest_relax_type; /* relax type on the coarsest grid */
    int         interp_type;        /* interpolation */
    INT         num_relax;          /* number of sweep */

} BMAMG_PARS;

typedef struct BMAMG_DATA_
{
    BMAMG_PARS          pars;

    HYPRE_IJMatrix      A;
    HYPRE_IJVector      b;
    HYPRE_IJVector      x;
    HYPRE_Solver        hypre_solver;
    INT                 ilower;
    INT                 iupper;

    BOOLEAN             assembled;

} BMAMG_DATA;
```

Figure 33: Data structure of AMG method

### 4.2.1 CPR-like Preconditioners

Linear systems from black oil and composition models are hard to solve, especially when the reservoirs are heterogeneous. However, the matrices from the pressure unknowns are positive definite, which can be solved by AMG methods. Many preconditioners have been developed to speed the solution of linear systems, such as the constrained pressure residual (CPR) method and FASP method [18]. Here we introduce our multi-stage preconditioners for the black oil and compositional models, which are based on the classical CPR method.

Numerical methods for black oil and compositional models may choose different unknowns [31]. Here we assume that the oil phase pressure $p_o$ is always one of the unknowns. The other variables are denoted as $\vec{s_i}$. The well unknowns are denoted by $\vec{w}$, whose dimension

```
static BMAMG_PARS amg_pars_default =
{
    /* maxit */              1,
    /* num_funcs */          -1,
    /* max_levels */         8,
    /* strength */           0.5,
    /* max_row_sum */        0.9,
    /* trunc error */        1e-2,

    /* coarsen_type */       Falgout,
    /* cycle_type */         v-cycle,
    /* relax_type */         hybrid Gauss-Seidel-forward,
    /* coarsest_relax_type */ hybrid symmetric Gauss-Seidel,
    /* interp type */        cmi,
    /* itr relax */          2,
};
```

Figure 34: Default parameters of AMG method

equals the number of wells in the reservoir, $N_w$. Let us define the pressure vector $p$ as:

$$p = \begin{pmatrix} p_{o,1} \\ p_{o,2} \\ \ldots \\ p_{o,N_g} \end{pmatrix}, \tag{5}$$

and the global unknown vector $x$ as:

$$x = \begin{pmatrix} p_{o,1} \\ p_{o,2} \\ \ldots \\ p_{o,N_g} \\ \vec{s_1} \\ \ldots \\ \vec{s_{N_g}} \\ \vec{w} \end{pmatrix}. \tag{6}$$

A restriction operator from $x$ to $p$ is defined as

$$\Pi_r x = p. \tag{7}$$

A prolongation operator $\Pi_p$ is defined as

$$\Pi_p p = \begin{pmatrix} p \\ \vec{0} \end{pmatrix}, \tag{8}$$

where $\Pi_p p$ has the same dimension as $x$.

If a proper ordering technique is applied, the matrix $A$ from black oil and compositional models can be written as equation (9),

$$A = \begin{pmatrix} A_{pp} & A_{ps} & A_{pw} \\ A_{sp} & A_{ss} & A_{sw} \\ A_{wp} & A_{ws} & A_{ww} \end{pmatrix}, \qquad (9)$$

where the sub-matrix $A_{pp}$ is the matrix corresponding to the pressure unknowns, the sub-matrix $A_{ss}$ is the matrix corresponding to the other unknowns, the sub-matrix $A_{ww}$ is the matrix corresponding to the well bottom hole pressure unknowns, and other matrices are coupled items.

Let us introduce some notations for the preconditioning system $Ax = f$. If $A$ is a positive definite matrix, then we define the notation $\mathscr{M}_g(A)^{-1}b$ to represent the solution $x$ from AMG methods. If it is solved by the RAS method, then we use the notation $\mathscr{R}(A)^{-1}b$ to represent solution $x$. The CPR-like preconditioners we develop are shown by Algorithm 5 to Algorithm 8, which are noted as CPR-FP, CPR-PF, CPR-FPF and CPR-FFPF methods [38], respectively.

---
**Algorithm 5** The CPR-FP preconditioner
---
1: $x = \mathscr{R}(A)^{-1}f$.
2: $r = f - Ax$
3: $x = x + \Pi_p(\mathscr{M}_g(A_{pp})^{-1}(\Pi_r r))$.
---

---
**Algorithm 6** The CPR-PF preconditioner
---
1: $x = \Pi_p(\mathscr{M}_g(A_{pp})^{-1}(\Pi_r f))$
2: $x = \mathscr{R}(A)^{-1}f$.
3: $r = f - Ax$
4: $x = x + \mathscr{R}(A)^{-1}f$.
---

---
**Algorithm 7** The CPR-FPF preconditioner
---
1: $x = \mathscr{R}(A)^{-1}f$.
2: $r = f - Ax$
3: $x = x + \Pi_p(\mathscr{M}_g(A_{pp})^{-1}(\Pi_r r))$.
4: $r = f - Ax$
5: $x = x + \mathscr{R}(A)^{-1}r$.
---

The data structure of the CPR preconditioners is shown in Figure 35. It has a RAS solver (`ras`) an AMG solver (`amg`), prolongation information (`pro_pres`) and `num_pro_pres`), buffers and settings for RAS solver and AMG solver. The term `CPR_PARS` stores parameters of the CPR methods.

**Algorithm 8** The CPR-FFPF preconditioner

1: $x = \mathscr{R}(A)^{-1}f$.
2: $r = f - Ax$
3: $x = x + \mathscr{R}(A)^{-1}r$.
4: $r = f - Ax$
5: $x = x + \Pi_p(\mathscr{M}_g(A_{pp})^{-1}(\Pi_r r))$.
6: $r = f - Ax$
7: $x = x + \mathscr{R}(A)^{-1}r$.

```
typedef struct CPR_PARS_
{
    RAS_PARS    ras;
    BMAMG_PARS  amg;
    INT         pres_which;
    INT         pres_loc;

    INT         itr_ras_pre;
    INT         itr_ras_post;

} CPR_PARS;

typedef struct CPR_DATA_
{
    RAS_DATA        ras;
    BMAMG_DATA      amg;

    INT             *pro_pres;   /* prolongation from pressure */
    INT             num_pro_pres;
    VEC             *varbuf;     /* vector r (A), buffer */
    VEC             *vpbbuf;     /* buffer for AMG */
    VEC             *vpxbuf;     /* buffer for AMG */

    CPR_PARS        pars;

} CPR_DATA;
```

Figure 35: Data structure of CPR preconditioners

# 5 Numerical Experiments

A Blue Gene/Q from IBM that located in the IBM Thomas J. Watson Research Center is employed. The system uses PowerPC A2 processor. Each processor has 18 cores and 16 cores are used for computation. Performance of each core is really low compared with processors from Intel. However, it has a strong network relative to compute performance and the system is scalable.

## 5.1 SpMV

**Example 1** *This example tests the performance of sparse matrix-vector multiplication (SpMV) on IBM Blue Gene/Q. The matrix is a square matrix from a Poisson equation and it has 200 millions rows. Its performance is shown in Table 1 and its scalability is presented in Figure 36.*

| # procs | 32 | 64 | 128 | 256 | 512 | 1024 |
|---|---|---|---|---|---|---|
| Time (s) | 2.211 | 1.078 | 0.556 | 0.269 | 0.134 | 0.067 |

Table 1: Performance of SpMV, Example 1

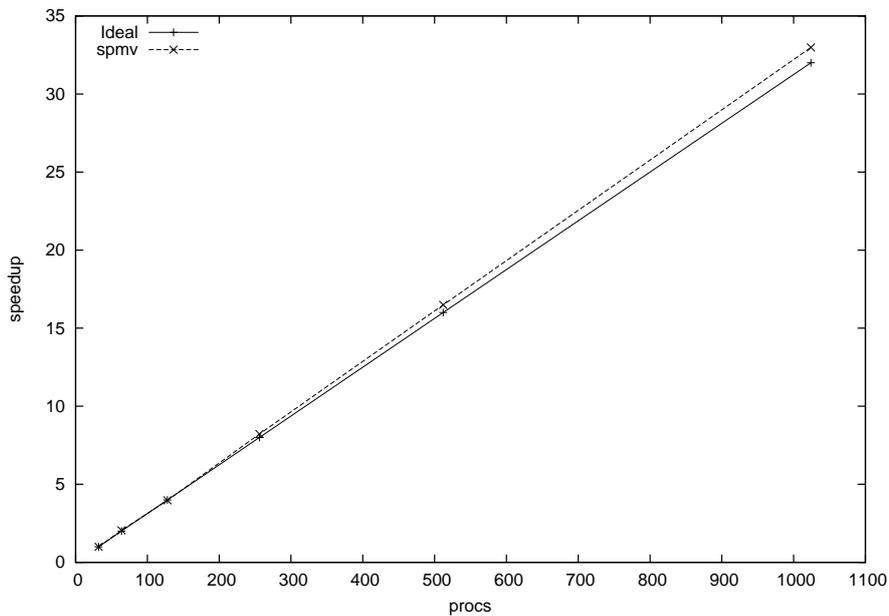

Figure 36: Scalability of SpMV, Example 1

This example uses up to 128 compute cards, when more than 128 MPI tasks are used, multiple MPI tasks run on one card. From Table 1, we can see that when MPI tasks are doubled, the running time of SpMV is reduced by half. This example show our SpMV kernel has excellent scalability. Speedup is compared with ideal condition in Figure 36, which shows that our SpMV kernel has good scalability.

## 5.2 Oil-water Model

**Example 2** *This example tests a refined SPE10 case for the two-phase oil-water model, where each grid cell is refined into 27 grid cells. This case has around 30 millions of grid cells and around 60 millions of unknowns. The stopping criterion for the inexact Newton method is 1e-3 and the maximal Newton iterations are 20. The BiCGSTAB solver is applied and its maximal iterations are 100. The preconditioner is the CPR-FPF preconditioner. The potential reordering and the Quasi-IMPES decoupling strategy are applied. The simulation* *period is 10 days. Up to 128 compute cards are used. The numerical summaries are shown in Table 2, and the speedup (scalability) is shown in Figure 37.*

| # procs | # Steps | # Ntn | # Slv | # Avg-S | Time (s) | Avg-T (s) |
|---|---|---|---|---|---|---|
| 64   | 50 | 315 | 3451 | 10.9 | 119167.4 | 378.3 |
| 128  | 48 | 286 | 3296 | 11.5 | 49488.7  | 173.0 |
| 256  | 54 | 323 | 4190 | 12.9 | 30423.2  | 94.1  |
| 512  | 52 | 329 | 3635 | 11.0 | 14276.5  | 43.3  |
| 1024 | 54 | 316 | 3969 | 12.5 | 7643.9   | 24.1  |

Table 2: Numerical summaries of Example 2

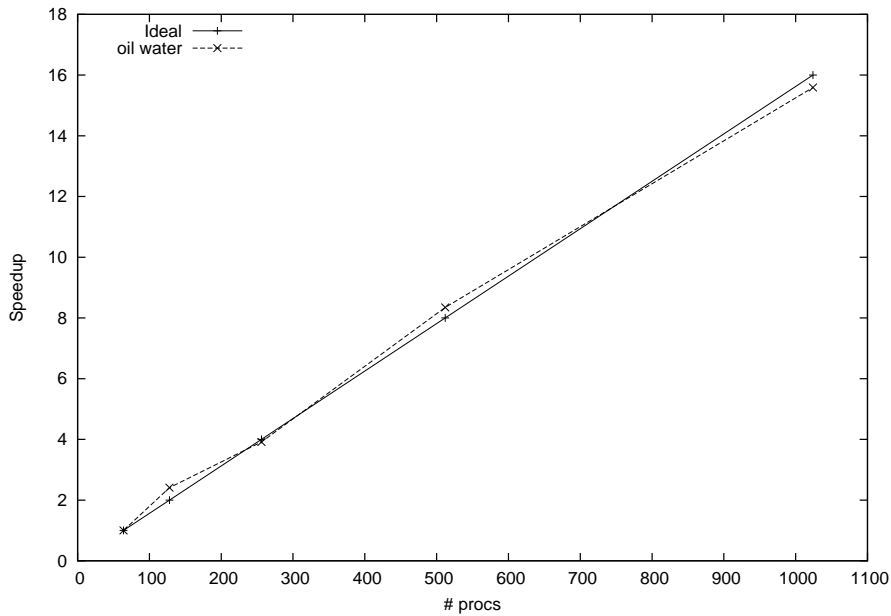

Figure 37: Scalability of Example 2

In this example, up to 1,024 MPI tasks are employed and the simulation with 64 MPI tasks is used as the base case to calculate speedup and scalability. The numerical summaries in Table 2 show the inexact Newton method is robust, where around 50 time steps and around 300 Newton iterations are used for each simulation with different MPI tasks. The linear solver BiCGSTAB and the preconditioner CPR-FPF show good convergence, where the average number of linear iterations for each nonlinear iteration is between 10 and 13. The results mean our linear solver and preconditioner are effective and efficient. The overall running time and average time for each Newton iteration show our simulator has excellent scalability on IBM Blue Gene/Q, which is almost ideal for parallel computing. The scalability is also demonstrated in Figure 37. The running time and scalability curve also demonstrate our linear solver and preconditioner are scalable for large-scale simulation.

**Example 3** *This example tests a refined SPE10 case for the two-phase oil-water model, where each grid cell is refined into 125 grid cells. This case has around 140 millions of grid cells and around 280 millions of unknowns. The stopping criterion for the inexact Newton method is 1e-2 and the maximal Newton iterations are 20. The GMRES(30) solver is applied and its maximal iterations are 100. The preconditioner is the CPR-FPF preconditioner. The potential reordering and the Quasi-IMPES decoupling strategy are applied. The simulation period is 2 days. Up to 128 compute cards are used. The numerical summaries are shown in Table 3, and the speedup (scalability) is shown in Figure 38.*

| # procs | # Steps | # Ntn | # Slv | # Avg-S | Time (s) | Avg-T (s) |
|---|---|---|---|---|---|---|
| 256  | 36 | 225 | 5544 | 24.6 | 117288.1 | 521.2 |
| 512  | 36 | 226 | 5724 | 25.3 | 57643.1  | 255.0 |
| 1024 | 35 | 207 | 5446 | 26.3 | 27370.3  | 132.2 |
| 2048 | 36 | 209 | 5530 | 26.4 | 14274.9  | 68.3  |

Table 3: Numerical summaries of Example 3

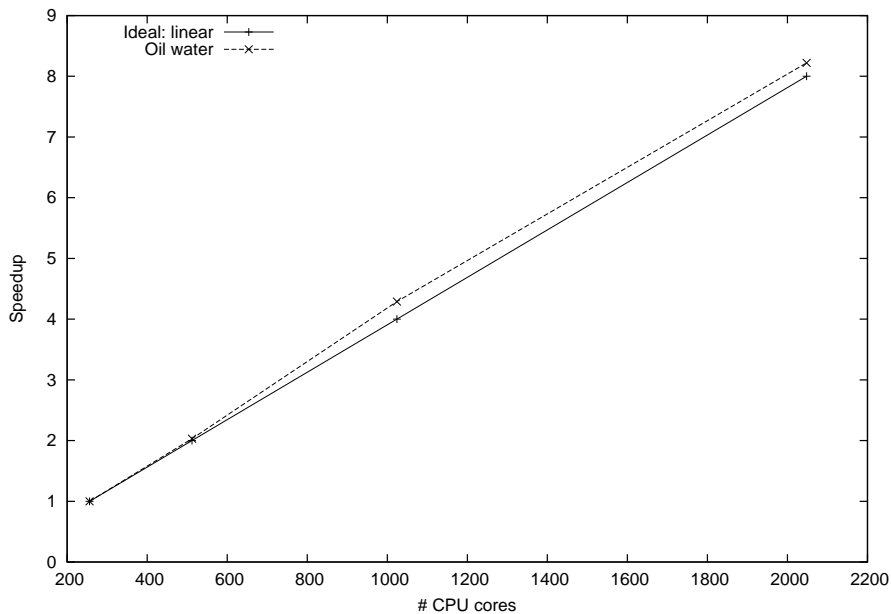

Figure 38: Scalability of Example 3

In this example, up to 2048 MPI tasks are employed and the simulation with 256 MPI tasks is used as the base case to calculate speedup and scalability. The numerical summaries in Table 3 show the inexact Newton method is robust, where around 36 time steps and around 220 Newton iterations are used for each simulation with different MPI tasks. The linear solver GMRES(30) and the preconditioner CPR-FPF show good convergence, where the average number of linear iterations for each nonlinear iteration is around 26. The overall running time and average time for each Newton iteration show our simulator has excellent scalability on IBM Blue Gene/Q, which is almost ideal for parallel computing and shows slight super-linear scalability. The scalability is also demonstrated in Figure 38. The results show our linear solver and preconditioner are scalable for large-scale simulation.

## 5.3 Black Oil Model

**Example 4** *The example tests the scalability of the black oil simulator using a refined SPE10 geological model, where each grid cell is refined to 27 grid cells. The model has 30.3 millions of grid cells. The inexact Newton method is applied and the termination tolerance is $10^{-2}$. The linear solver is BiCGSTAB, whose maximal inner iterations are 100. The preconditioner is the CPR-FPF method and the overlap for the RAS method is one. The potential reordering and the ABF methods are enabled. The simulation period is 10 days. The maximal change allowed in one time step of pressure is 1,000 psi and the maximal change of saturation is 0.2. Up to 128 compute cards are used. Summaries of numerical results are shown in Table 4.*

| # procs | # Steps | # Ntn | # Slv | # Avg-S | Time (s) |
|---|---|---|---|---|---|
| 64   | 33 | 292 | 1185 | 4.0 | 106265.9 |
| 128  | 33 | 296 | 1150 | 3.8 | 50148.3 |
| 256  | 33 | 299 | 1267 | 4.2 | 25395.8 |
| 512  | 33 | 301 | 1149 | 3.8 | 12720.5 |
| 1024 | 33 | 301 | 1145 | 3.8 | 6814.2 |

Table 4: Numerical summaries of Example 4

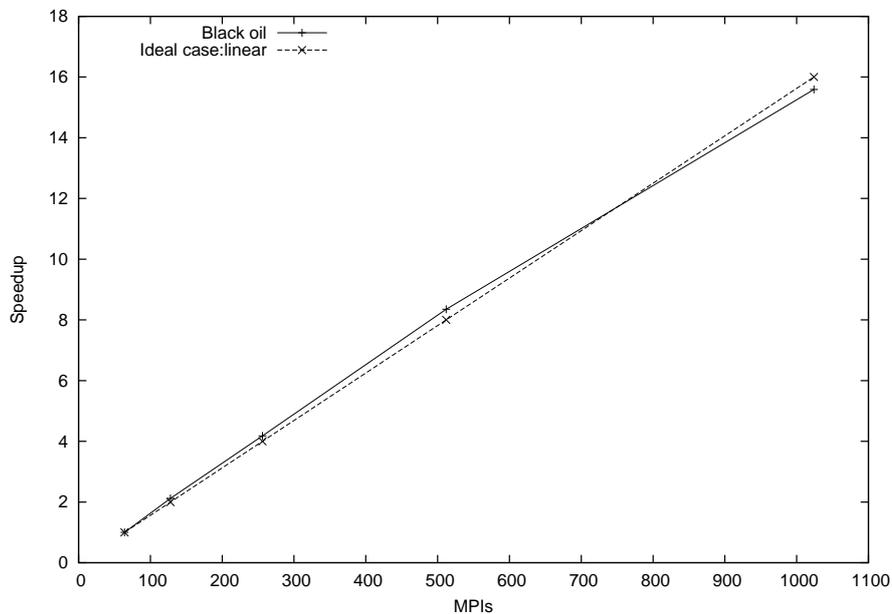

Figure 39: Scalability (speedup) of Example 4

Table 4 includes information for the nonlinear method, linear solver and running time. For all simulations, 33 time steps are used and the total Newton iterations are around 300. The results show the inexact Newton method is robust. For the linear solver and preconditioner, their convergence is good, which terminate in around 4 iterations. The results mean the linear solver and preconditioner are robust and effective for this highly heterogeneous model. The running time, average time per Newton iteration and scalability curve in Figure 39 show the scalability of our simulator, linear solver and preconditioner is good. When we use up to 1,024 MPI tasks and each compute card runs up to 8 MPI tasks, the scalability is excellent.

**Example 5** *The case is a refined SPE1 project with 100 millions of grid cells. Linear solver is BiCGSTAB. Potential reordering and ABF decoupling are applied. Numerical summaries are in Table 5.*

| MPIs | # Steps | # Newton | # Solver | # Avg. Itr | Time (s) |
|---|---|---|---|---|---|
| 512  | 27 | 140 | 586 | 4.1 | 11827.9 |
| 1024 | 27 | 129 | 377 | 2.9 | 5328.4  |
| 2048 | 26 | 122 | 362 | 2.9 | 2708.5  |
| 4096 | 27 | 129 | 394 | 3.0 | 1474.2  |

Table 5: Numerical summaries, Example 5

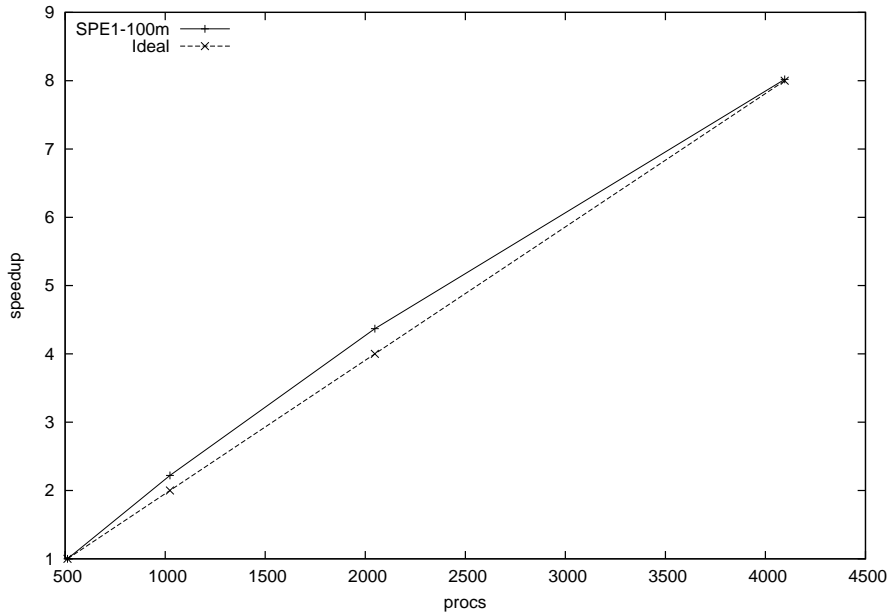

Figure 40: Scalability of Example 5

## 5.4 Poisson Equation

**Example 6** *This example tests a Poisson equation with 3 billions of grid cells. The example uses up to 4,096 CPU cores (MPIs). The linear solver is GMRES(30) method with RAS preconditioner. The solver runs 90 iterations. The overlap of RAS method is one, and subdomain problem on each core is solved by ILU(0). The numerical summaries are reported in Table 6 and scalability results are shown in Figure 41.*

| # procs | Gridding | Building | Assemble | Overall (s) | Speedup |
|---------|----------|----------|----------|-------------|---------|
| 512     | 217.0    | 29.16    | 66.42    | 918.91      | 1.0     |
| 1024    | 98.83    | 14.79    | 33.71    | 454.04      | 2.02    |
| 2048    | 47.53    | 7.49     | 17.47    | 227.05      | 4.05    |
| 4096    | 23.31    | 3.86     | 9.17     | 116.64      | 7.88    |

Table 6: Numerical summaries of Example 6

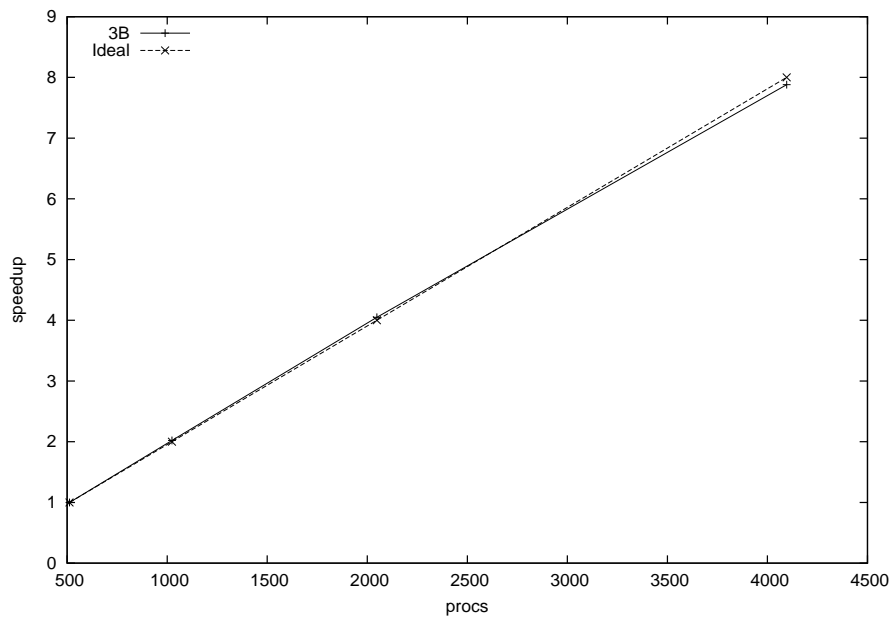

Figure 41: Scalability of Example 6

Table 6 shows numerical summaries of Poisson equation. This example tests strong scalability. Building time is the time spent on generation of a linear system $Ax = b$. Since there is no communication involved, the scalability is ideal. Assemble time includes time for generating linear solver, and time for generating preconditioner (RAS method). The overall time includes gridding time, building time, assemble time and solution time. Again, from Table 6 and Figure 41, we can see our solvers have excellent scalability.

**Example 7** *The case goes with the size of the matrix at 6 billion variables and is constructed from pressure equations. GMRES(30) was applied to solve the system with RAS (Restricted Additive Schwarz) as the preconditioner and fixed iterations at 90. IBM Blue Gene/Q was* *used to carry the simulation. Numerical summaries are shown in Table 7 and scalability is presented in Fig 42 [39].*

Table 7: Numerical summaries of the large-scale model

| # procs | Gridding (s) | Build (s) | Assemble (s) | Overall (s) |
|---------|--------------|-----------|--------------|-------------|
| 512     | 429.78       | 57.83     | 130.54       | 1829.18     |
| 1024    | 200.24       | 29.28     | 66.12        | 906.73      |
| 2048    | 106.83       | 14.83     | 34.03        | 463.59      |
| 4096    | 47.82        | 7.59      | 17.7         | 232.77      |

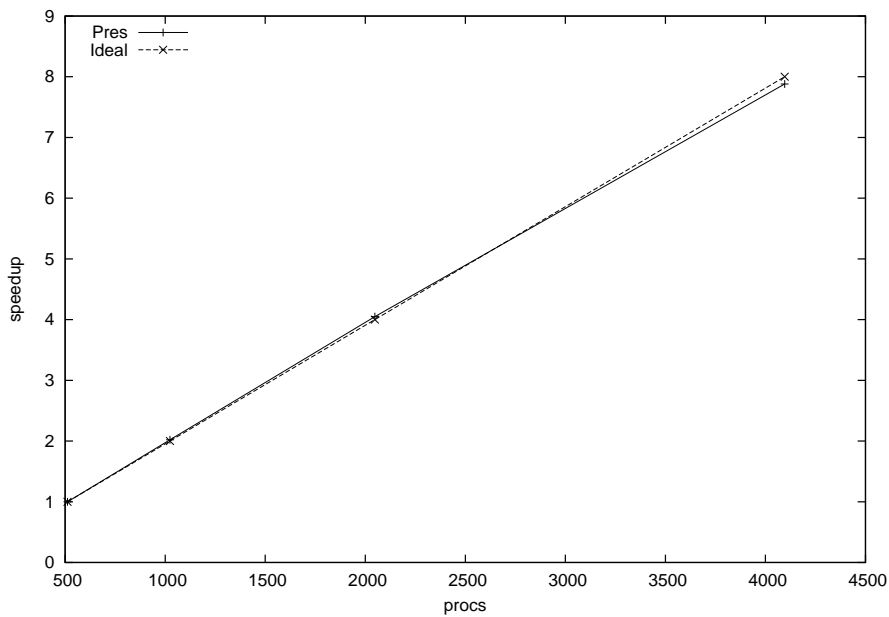

Figure 42: Scalability of the large-scale model

This example tests the scalability of grid generation, building of linear system, and solution of linear system (including solver, preconditioner and SpMV). Table 7 shows that when MPI tasks are doubled, running time of grid generation, building of linear system, and solution of linear system is cut by half, which means our platform and linear solvers have excellent scalability. The scalability is demonstrated by Figure 42 demonstrate. This example also show that the solver can calculate extremely large-scale linear systems.

# 6  Conclusion

Our work on developing in-house parallel linear solvers is presented in this paper, which implements solvers and preconditioners for reservoir simulators. Various techniques and data structures have been introduced, including distributed matrices and vectors, and multi-state preconditioners for reservoir simulations. Numerical results show that our solvers have excellent scalability and applications based on the solvers can be sped up thousands of times faster.

# Acknowledgement

The support of Department of Chemical and Petroleum Engineering and Reservoir Simulation Group, University of Calgary is gratefully acknowledged. The research is partly supported by NSERC/AIEES/Foundation CMG and AITF Chairs.